# Neutron activation analysis of the $^{30}$Si content of highly enriched $^{28}$Si: proof of concept and estimation of the achievable uncertainty

G. D'Agostino(1), G. Mana(2), M. Oddone(3), M. Prata(4), L. Bergamaschi(1) and L. Giordani(1)

(1) Istituto Nazionale di Ricerca Metrologica (INRIM), Unit of Radiochemistry and Spectroscopy, c/o Department of Chemistry, University of Pavia, via Taramelli 12, 27100 Pavia, Italy

(2) Istituto Nazionale di Ricerca Metrologica (INRIM), Strada delle Cacce 91, 10135 Torino, Italy.

(3) Department of Chemistry, University of Pavia, via Taramelli 12, 27100 Pavia, Italy

(4) Laboratorio Energia Nucleare Applicata (LENA), University of Pavia, via Aselli 41, 27100 Pavia, Italy

Email: g.dagostino@inrim.it

**Abstract:** We investigated the use of neutron activation to estimate the $^{30}$Si mole fraction of the ultra-pure silicon material highly enriched in $^{28}$Si for the measurement of the Avogadro constant. Specifically, we developed a relative method based on Instrumental Neutron Activation Analysis and using a natural-Si sample as a standard. To evaluate the achievable uncertainty, we irradiated a 6 g sample of a natural-Si material and modeled experimentally the signal that would be produced by a sample of the $^{28}$Si-enriched material of similar mass and subjected to the same measurement conditions. The extrapolation of the expected uncertainty from the experimental data indicates that a measurement of the $^{30}$Si mole fraction of the $^{28}$Si-enriched material might reach a 4% relative combined standard uncertainty.

## 1. Introduction

The measurement of the Avogadro constant, $N_A$, requires the knowledge of the molar mass, $M$, and hence of the isotopic composition, of ultra-pure silicon materials [1-3]. At present, the technique with the lowest uncertainty of measurement for the assessment of the silicon molar mass is the Isotope Dilution Mass Spectrometry (IDMS) [4]. However, when using natural-Si, the $5 \times 10^{-6}$ $M$ combined standard uncertainty of the molar mass measurement [5] prevents the targeted $1 \times 10^{-8}$ $N_A$ combined standard uncertainty to be achieved [1]. The use of silicon highly enriched in $^{28}$Si overcomes this barrier and opened the way to reach the necessary uncertainty: the relative combined standard uncertainty of the $^{28}$Si-enriched molar mass was reduced to below the $10^{-8}$ level [6-10].

Several samples from parts 5 and 8 of the Avogadro $^{28}$Si-enriched crystal AVO28 [11] have been measured by IDMS. The $x(^{28}\text{Si})$, $x(^{29}\text{Si})$ and $x(^{30}\text{Si})$ mole fraction values obtained by the Physikalish-Technische Bundesanstalt (PTB), the National Research Council of Canada (NRC), the National Metrology Institute of Japan (NMIJ) and the National Institute of Standards and Technology (NIST) are given in table 1. With the exception of the NRC datum, there is a general agreement among the measurement results.

| Lab. | $x(^{28}Si)/(mol\ mol^{-1})$ | $x(^{29}Si)/(mol\ mol^{-1})$ | $x(^{30}Si)/(mol\ mol^{-1})$ | AVO28 parts |
|---|---|---|---|---|
| PTB [6] | 0.999 957 50(17) | 0.000 041 21(15) | 0.000 001 290(40) | 5, 8 |
| PTB [7] | 0.999 957 26(17) | 0.000 041 62(17) | 0.000 001 120(06) | 5, 8 |
| NRC [8] | 0.999 958 79(19) | 0.000 040 54(14) | 0.000 000 670(60) | 5, 8 |
| NIMIJ [9] | 0.999 957 63(07) | 0.000 041 20(14) | 0.000 001 180(69) | 5, 8 |
| NIST [10] | 0.999 957 703(38) | 0.000 041 229(37) | 0.000 001 069(06) | 5, 8 |

**Table 1.** Measured isotopic compositions of the AVO28 material with their associated combined standard uncertainties.

The $^{29}Si$ and $^{30}Si$ mole fractions are plotted in figure 1. The PTB, NMIJ and NIST results show a difference in $x(^{29}Si)$ and $x(^{30}Si)$ relative to their average lower than 1% and 11%, respectively. In contrast, the $x(^{29}Si)$ and $x(^{30}Si)$ values obtained by NRC are significantly smaller than the average of the values obtained by the other three laboratories.

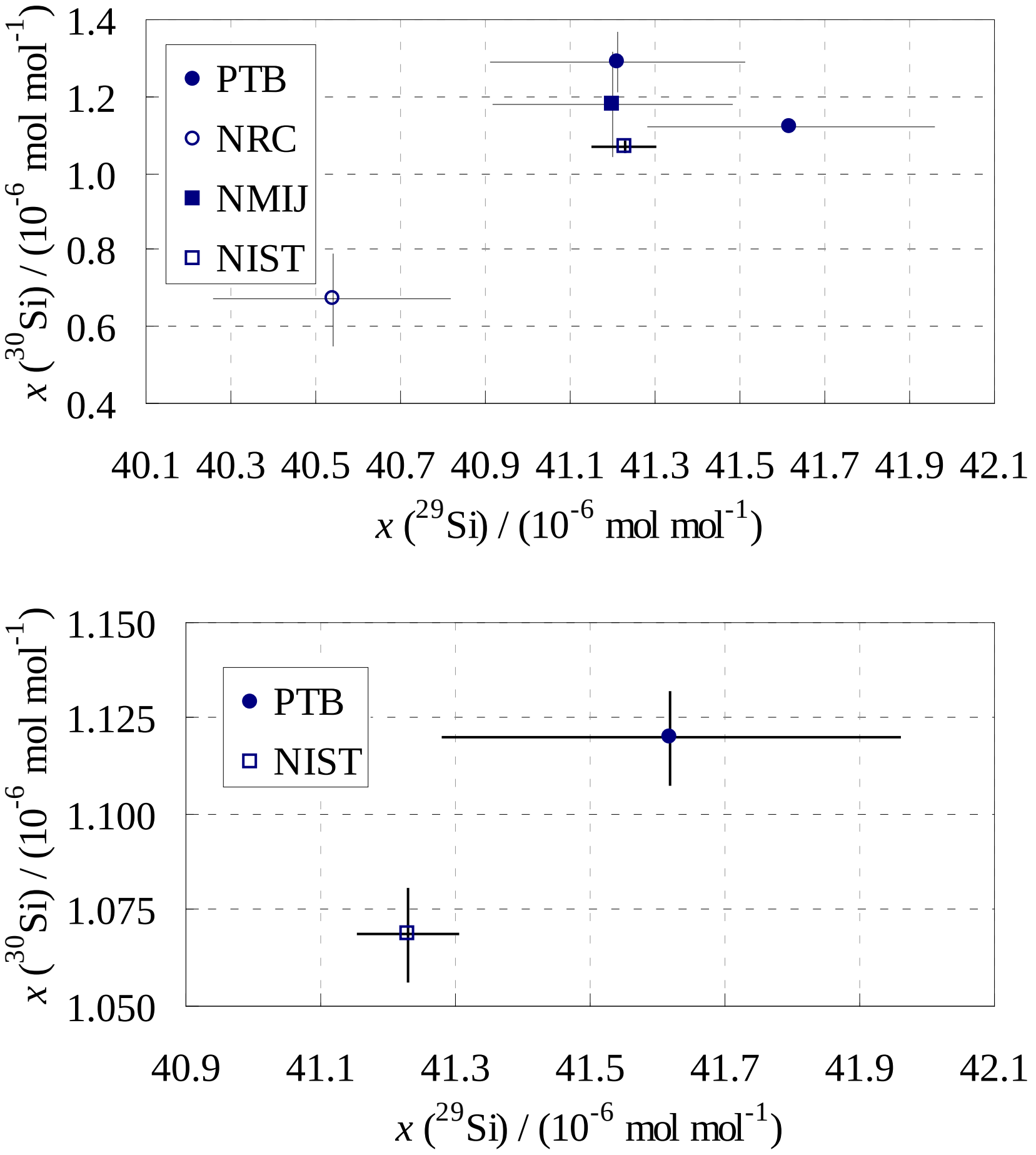

**Figure 1.** Overall view of the measured $x(^{29}Si)$ and $x(^{30}Si)$ values of the AVO28 material (upper graph) and a magnification of the PTB and NIST values (lower graph); the bars indicate a coverage factor 2.

Aiming at collecting experimental data with a different technique and at investigating the discrepancy of the NRC data, we propose to measure $x(^{30}Si)$ by using Instrumental Neutron Activation Analysis (INAA). When designing the measurement, particular care must be given to the $10^{-6}$ mol/mol level of $^{30}Si$ in the AVO28 material (see table 1 and figure 1).

Being the $x(^{30}Si)$ measurement demanding and owing to the limited availability of samples of the AVO28 material, we carried out a feasibility study using natural silicon. Specifically, we neutron irradiated a natural-Si sample and measured the $\gamma$-emission of $^{31}Si$ during its radioactive decay. These data were used to refine the estimate of the half-life of $^{31}Si$ and to model experimentally the $\gamma$-emission of $^{31}Si$ that would be produced in a similar AVO28 sample when applying the same irradiation and counting conditions. Eventually, we estimated the expected relative combined standard uncertainty of the measurement.

## 2. Measurement method

The neutron activation analysis is based on the exposure of the samples to a neutron flux and the detection of the $\gamma$ rays emitted by the produced radionuclides. In particular, during the irradiation, the stable nuclei within the samples are transformed into radioactive nuclei by neutron capture reactions. After irradiation, the activated nuclei decay by emitting $\gamma$ rays with specific energies. Since the amount of produced radionuclides depends on the amount of target nuclides in the samples, the latter are quantified by measuring the intensity of the emitted $\gamma$ rays.

### *2.1. Measurement equation*

Owing to the high material purity, the $^{30}Si$ mole fraction of a AVO28 sample is

$$x(^{30}\mathrm{Si}_{\mathrm{AVO28}}) = \frac{N(^{30}\mathrm{Si}_{\mathrm{AVO28}})}{N(\mathrm{Si}_{\mathrm{AVO28}})}, \tag{1}$$

where $N(^{30}\mathrm{Si}_{\mathrm{AVO28}})$ is the number of $^{30}Si$ atoms and $N(\mathrm{Si}_{\mathrm{AVO28}})$ is the total number of Si atoms.

By using a ultra-pure natural-Si sample of known isotopic composition as a standard, we can measure $x(^{30}\mathrm{Si}_{\mathrm{AVO28}})$ according to

$$x(^{30}\mathrm{Si}_{\mathrm{AVO28}}) = k\, x(^{30}\mathrm{Si}_{\mathrm{STD}}) \frac{m_{\mathrm{STD}}}{m_{\mathrm{AVO28}}} \frac{A_{\mathrm{r\,AVO28}}}{A_{\mathrm{r\,STD}}}, \tag{2}$$

where $x(^{30}\mathrm{Si}_{\mathrm{STD}})$ is the $^{30}Si$ mole fraction of the standard, $m_{\mathrm{STD}}$ and $m_{\mathrm{AVO28}}$ are the masses of the standard and AVO28 samples, $A_{\mathrm{r\,STD}}$ and $A_{\mathrm{r\,AVO28}}$ are the relative atomic weights of Si in the standard and AVO28 materials, respectively, $k = N(^{30}\mathrm{Si}_{\mathrm{AVO28}})/N(^{30}\mathrm{Si}_{\mathrm{STD}})$, and $N(^{30}\mathrm{Si}_{\mathrm{STD}})$ is the number of $^{30}Si$ atoms in the standard.

The measurement of $k$ is central to this proposal and can be carried out in a neutron activation experiment by counting of the 1266.1 keV $\gamma$-photons emitted by the standard and AVO28 samples during the radioactive decay of $^{31}Si$. The $^{31}Si$, which has an half-life of about 2.6 h, is produced by activation of $^{30}Si$ via the neutron capture reaction $^{30}Si(n,\gamma)^{31}Si$. Additional neutron capture reactions occurring in silicon and producing $\gamma$-emitting radionuclides include $^{28}Si(n,p)^{28}Al$ and $^{29}Si(n,p)^{29}Al$. Because the half-lives of $^{28}Al$ and $^{29}Al$ are of the order of minutes, the counting of the $^{31}Si$ $\gamma$-photons becomes relatively free from matrix interferences after sufficient decay of the Al radionuclides. In addition, since the decay scheme of $^{31}Si$ shows only one energy transition, the samples can be counted in contact with the end-cap of the detector without the effect of true coincidence summing.

When co-irradiating the standard and AVO28 samples, the activation equation can be rearranged to compute the ratio $k$ according to

$$k = \frac{\left.\dfrac{C}{e^{-\lambda t_\mathrm{d}} k_\mathrm{ss}\, R\, k_\mathrm{sa}\, k_\mathrm{g}\, \varepsilon}\right|_\mathrm{AVO28}}{\left.\dfrac{C}{e^{-\lambda t_\mathrm{d}} k_\mathrm{ss}\, R\, k_\mathrm{sa}\, k_\mathrm{g}\, \varepsilon}\right|_\mathrm{STD}}, \tag{3}$$

where the subscripts STD and AVO28 refer to the standard and AVO28 samples, respectively, $R$ is the (n,γ) reaction rate per target nuclide (i.e. per $^{30}$Si atom), $\lambda$ is the decay constant of $^{31}$Si, $C$ is the count rate at a decay time $t_\mathrm{d}$, $k_\mathrm{ss}$, $k_\mathrm{sa}$, and $k_\mathrm{g}$ are the irradiation self-shielding, the emission self-absorption and the geometry factors, respectively, $\varepsilon$ is the detection full-energy $\gamma$ efficiency for a point-like source located at the center of mass of the sample. The (n,γ) reaction rate per $^{30}$Si atom is $R = \int \sigma(E)\, \varphi(E)\, dE$, where $E$ is the neutron energy, $\sigma(E)$ is the (n,γ) reaction cross section of $^{30}$Si and $\varphi(E)$ is the energy spectrum of the neutrons. The decay constant is $\lambda = \ln(2)/t_{1/2}$, where $t_{1/2}$ is the half-life of $^{31}$Si.

In equation (3), the count rates $C(t_\mathrm{d})$ are obtained by averaging several values extrapolated from two spectrometry sequences starting at $t_\mathrm{d\,STD}$ and $t_\mathrm{d\,AVO28}$ and consisting of consecutive counts carried out during the decay of $^{31}$Si. More specifically, each count rate value extrapolated from the $i$-th count of the sequence, starting at $t_i$ and lasting $t_\mathrm{c}$, is

$$C_i(t_\mathrm{d}) = \frac{\lambda\, n_\mathrm{c}}{e^{-\lambda (t_i - t_\mathrm{d})}(1 - e^{-\lambda t_\mathrm{c}})} \frac{t_\mathrm{c}}{t_\mathrm{c} - t_\mathrm{dead}}, \tag{4}$$

where the subscript STD or AVO28 has been omitted, $n_\mathrm{c}$ is the $i$-th total net count stored at the energy bins of the multichannel analyzer calibrated to collect the $\gamma$-photons at 1266.1 keV and $t_\mathrm{dead}$ is the dead time of the detection system during the $i$-th count. The equation (4) does not include the random summing factor because a pile-up rejection circuit makes this effect negligible at the expected low count rates.

According to the equations (2) and (3), the measurement equation of $x(^{30}\mathrm{Si}_\mathrm{AVO28})$ is

$$x(^{30}\mathrm{Si}_\mathrm{AVO28}) = \kappa_\mathrm{R}\, \kappa_\mathrm{td}\, \kappa_\varepsilon\, \kappa_\mathrm{ss}\, \kappa_\mathrm{sa}\, \kappa_\mathrm{g} \frac{C_\mathrm{AVO28}}{C_\mathrm{STD}} x(^{30}\mathrm{Si}_\mathrm{STD}) \frac{m_\mathrm{STD}}{m_\mathrm{AVO28}} \frac{A_\mathrm{r\,AVO28}}{A_\mathrm{r\,STD}}. \tag{5}$$

where we set $\kappa_\mathrm{R} = R_\mathrm{STD}/R_\mathrm{AVO28}$, $\kappa_\mathrm{td} = e^{-\lambda(t_\mathrm{d\,AVO28} - t_\mathrm{d\,STD})}$, $\kappa_\varepsilon = \varepsilon_\mathrm{STD}/\varepsilon_\mathrm{AVO28}$, $\kappa_\mathrm{ss} = k_\mathrm{ss\,STD}/k_\mathrm{ss\,AVO28}$, $\kappa_\mathrm{sa} = k_\mathrm{sa\,STD}/k_\mathrm{sa\,AVO28}$ and $\kappa_\mathrm{g} = k_\mathrm{g\,STD}/k_\mathrm{g\,AVO28}$. The correction factors $\kappa_\mathrm{R}$ and $\kappa_\mathrm{td}$ take the spatial variations of amplitude and distribution of the neutron energy spectrum at the irradiation positions and the different starting times $t_\mathrm{d}$ of the counting sequence into account.

*2.2. Measurement procedure*

The application of equation (5) is challenging because $x(^{30}\mathrm{Si}_{\mathrm{AVO28}})$ is at the $10^{-6}$ level. The $\gamma$-emission of $^{31}$Si produced in an AVO28 sample may not allow a good counting statistics. In the extreme case, the $\gamma$ count might be below the detection limit of the experiment. In an effort to decrease the detection limit, the AVO28 sample must be (i) shaped to have the maximum mass for the neutron irradiation, (ii) exposed to the highest accessible neutron flux until the production of $^{31}$Si is almost saturated, (iii) counted as soon as possible after irradiation and (iv) placed close to the end-cap of the detector to maximize the detection efficiency.

To reduce the uncertainty of $\kappa_{\mathrm{ss}}$, $\kappa_{\mathrm{sa}}$, $\kappa_{\mathrm{g}}$ and $\kappa_{\varepsilon}$, both the standard and AVO28 samples must be (i) cylinders with equal length and diameter and (ii) counted at the same position with respect to the detector. Additionally, since the $x(^{30}\mathrm{Si}_{\mathrm{STD}})$ is $3.1 \times 10^{-2}$ [12], the $^{31}$Si count rate of the AVO28 sample is $3.2 \times 10^{-5}$ times the $^{31}$Si count rate of the natural-Si standard. Therefore, to avoid the detector saturation, the spectrometry sequence of the standard must start several $^{31}$Si half-lives after the spectrometry sequence of the AVO28 sample has been completed.

## 3. Experimental

The target of the experimental test was to estimate the half-life of $^{31}$Si and to extrapolate the count rate of $^{31}$Si that is observed when irradiating a AVO28 sample. The sample used was a 6 g piece of the WASO04 material, a high-purity natural-Si crystal suitable to provide a standard for the AVO28 measurement [13]. Previous measurements confirmed that this material is substantially free of contaminant elements, at least those that can detected by INAA [14]. The shape and the size of the WASO04 sample, shown in figure 2, were of no concern because they don't affect the information being sought.

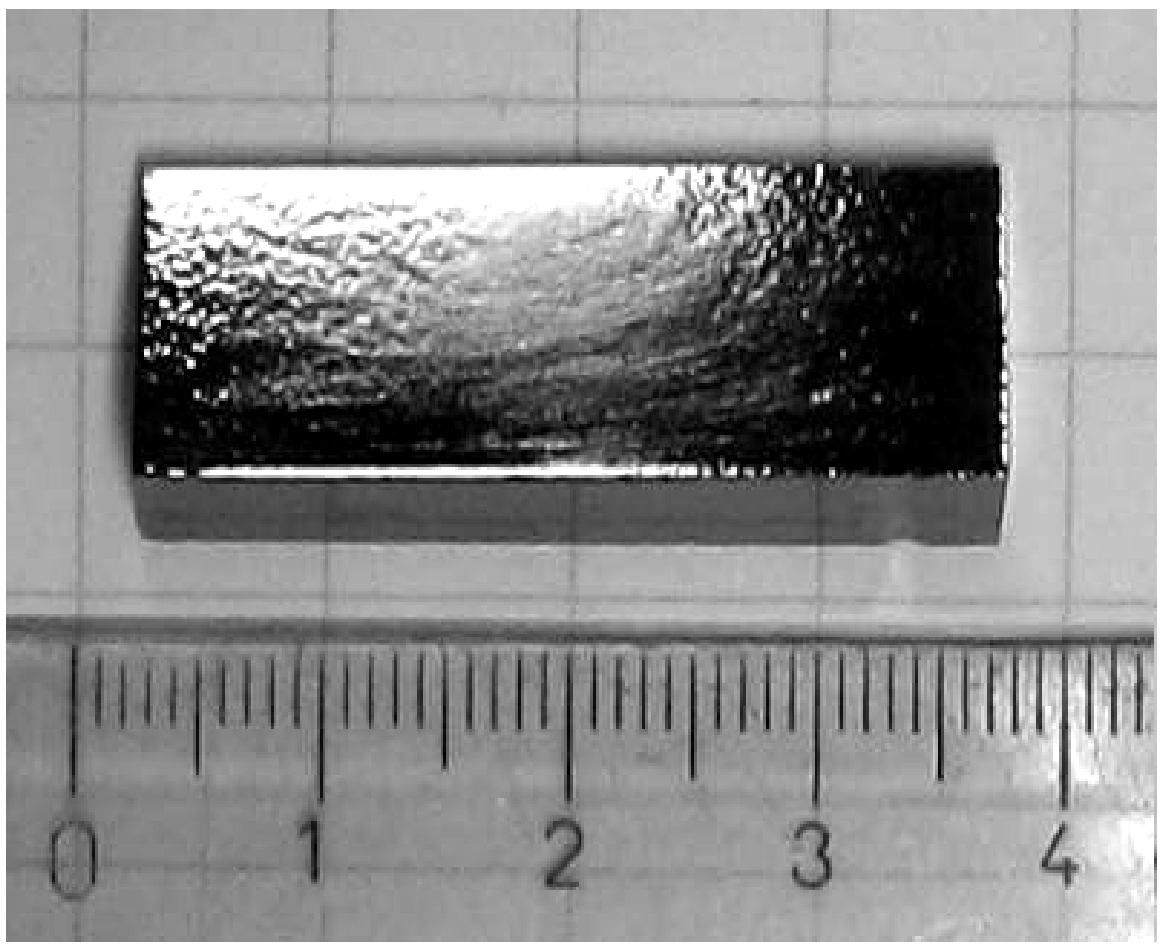

**Figure 2.** The WASO04 sample.

To minimize the interference signals produced by any surface contamination, the sample was strongly etched (solution 10:1, nitric acid and hydrofluoric acid) just before the neutron irradiation. The etching removed about 80 mg from the surface of the sample. The irradiated mass was measured with a calibrated digital analytical balance having a resolution of 10 µg. The sample was then sealed in a 8 mL polyethylene vial and inserted into a polyethylene container for the irradiation. The neutron irradiation lasted 3 h and took place in the central thimble of the 250 kW TRIGA Mark II reactor at the Laboratory of Applied Nuclear Energy (LENA) of the University of Pavia. The nominal thermal and epithermal neutron fluxes are $6 \times 10^{12}$ cm$^{-2}$ s$^{-1}$ and $5.5 \times 10^{11}$ cm$^{-2}$ s$^{-1}$, respectively.

At the end of the irradiation, the sample was left to cool for about 3 h, removed from the vial, rinsed with diluted nitric acid and placed in a polycarbonate container for counting. The loss of mass due to rinsing was negligible. The $\gamma$-photon detection was carried out by a coaxial Ge detector Ortec (relative efficiency 50%, resolution 1.90 keV FWHM at 1332 keV) connected to a digital signal processor Ortec DSPEC jr 2.0 and a dedicated computer. The data acquisition and processing were done by using the software GammaVision [15]. The energy and the FWHM calibrations were made by using a standard multi-$\gamma$ source manufactured by the Laboratoire Etalons d'Activité, type 9ML01EGMA.

*3.1. Half-life of $^{31}$Si*

The half-life of $^{31}$Si was estimated by measuring the decay of the count rate of the irradiated sample in two successive recording sequences. Each sequence consisted of several repeated counts of the 1266.1 keV $\gamma$-photons emitted by $^{31}$Si. The counting windows were adjusted on-line in such a way that the relative standard uncertainty due to countings was always 1%.

The first sequence started 3 h after the irradiation and lasted 17 h. The sample was placed about 26 cm from the end-cap of the detector. The second sequence started 27 h after the irradiation and lasted 10 h. The sample was placed about 1 cm from the end-cap of the detector. To limit the uncertainty due to the dead time of the detection system, the data collected with a relative dead time, $t_{\mathrm{dead}}/t_{\mathrm{c}}$ , higher than 30% were rejected.

Each collected net count, $n_{\mathrm{c}}$, was used to compute the detection count rate, $C(t_i)$, at the beginning of the $i$-th count according to

$$C(t_i) = \frac{\lambda\, n_{\mathrm{c}}}{(1 - e^{-\lambda t_{\mathrm{c}}})} \frac{t_{\mathrm{c}}}{t_{\mathrm{c}} - t_{\mathrm{dead}}} . \qquad (6)$$

The results are shown in figure 3.

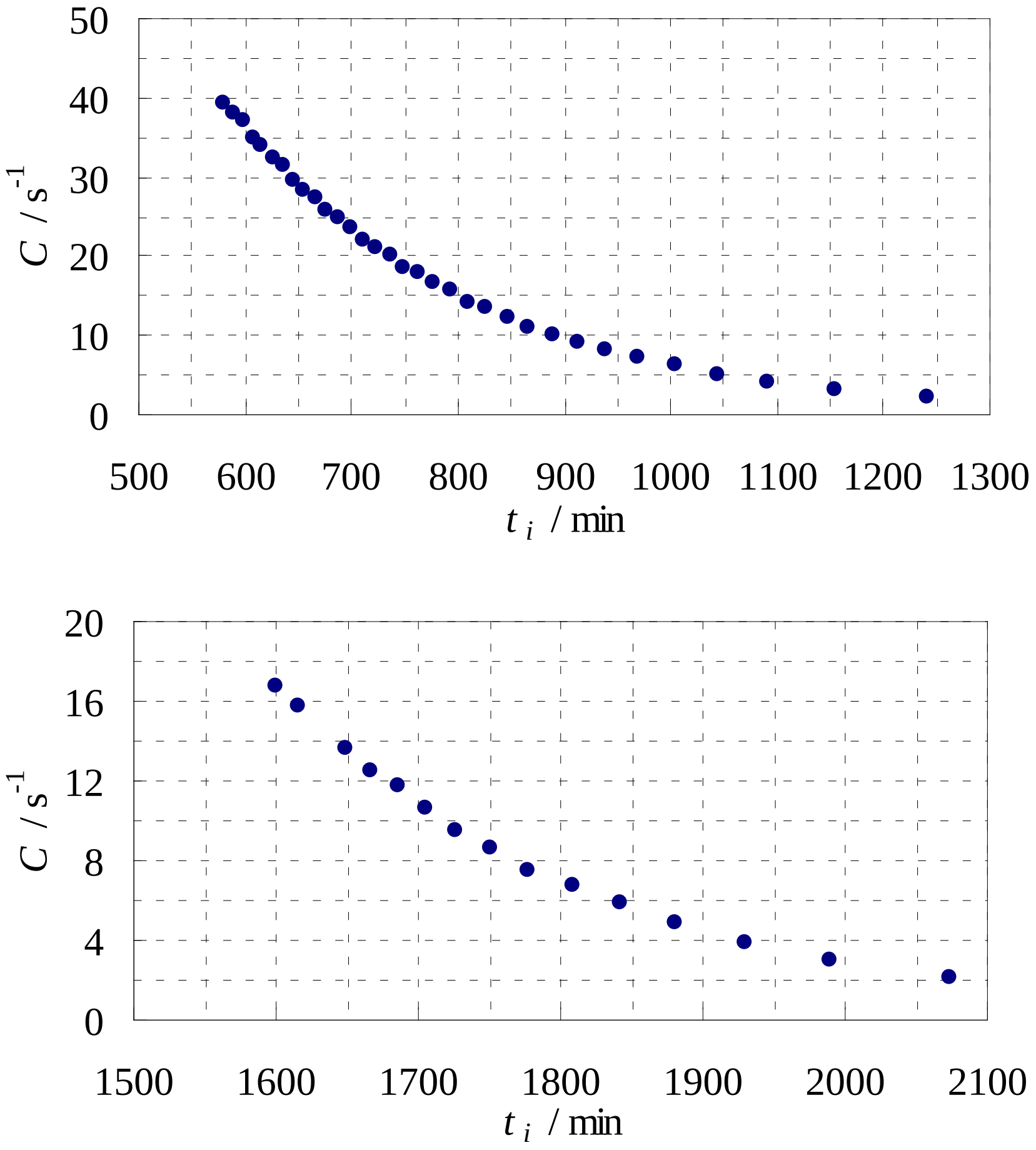

**Figure 3.** Count rates measured during the decay of $^{31}Si$ with the sample placed 26 cm (upper graph) and 1 cm (lower graph) from the detector end-cap; $t_i$ = 0 min corresponds to the irradiation end.

To estimate the half-life we fitted the model $y = a\,e^{-b\,x}$ to the experimental data. As a first step, the literature value of $t_{1/2}$ was used to compute the value of $\lambda$ in equation (6). Next, the value of $t_{1/2} = \ln(2)/b$ obtained by the fit was used to refine the $\lambda$ value and to re-compute the data according to equation (6). The fitting was iterated until the variation of $t_{1/2}$ was negligible. We used a non-linear curve fitting algorithm from the software OriginPro [16]. The fitted data were not weighted because the data acquisition was designed to have a constant relative standard uncertainty of 1%. The residuals are shown in figure 4.

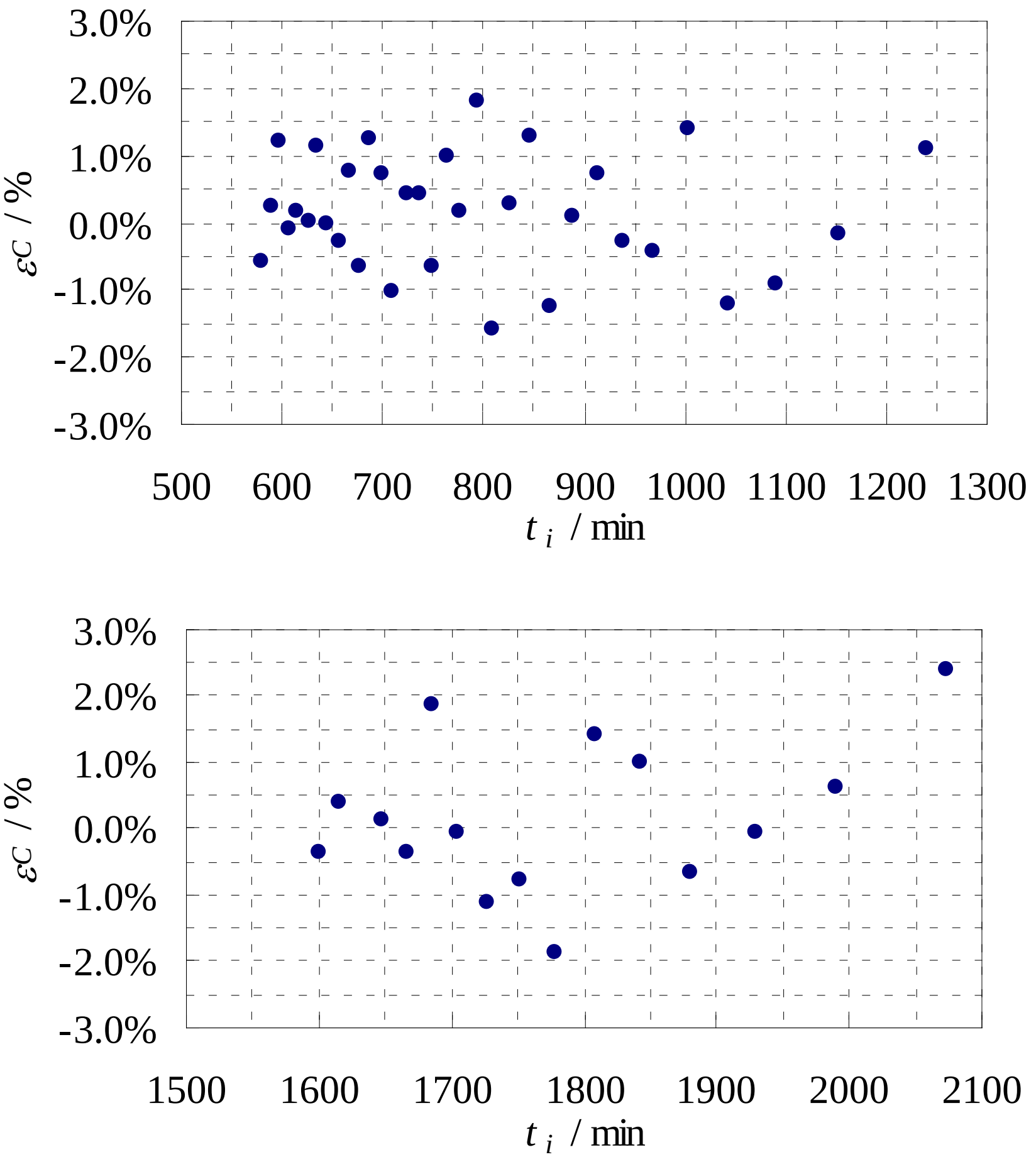

**Figure 4.** Relative residuals after fitting the count rates measured during the decay of $^{31}Si$ with the sample placed 26 cm (upper graph) and 1 cm (lower graph) from the detector end-cap; $t_i = 0$ min corresponds to the irradiation end.

The $t_{1/2}$ values obtained from the first and second sequence were 157.65(0.56) min and 156.76(1.07) min, respectively. The weighted average, 157.46(0.50) min, is consistent with the literature value, 157.36(0.26) min [17]. Additionally, the 1% standard deviation of the residuals endorses both the detection system and data processing.

### *3.2. Expected $^{31}Si$ count rate*

The collected data were used to model experimentally the $^{31}Si$ count rate of an AVO28 sample having the mass of the measured WASO04 piece. The AVO28 sample must be counted as soon as possible after irradiation and close to the detector. To estimate the expected count rate, we back-extrapolated at the end of the irradiation the WASO04 count rate observed during the second spectrometry sequence, where the WASO04 sample was placed 1 cm from the detector. The count rate value we obtained was about $20 \times 10^3$ $s^{-1}$. In addition, since the signal of the AVO28 sample is $3.2 \times 10^{-5}$ times the signal of the WASO04 sample, we rescaled the data accordingly. The result is shown in figure 5.

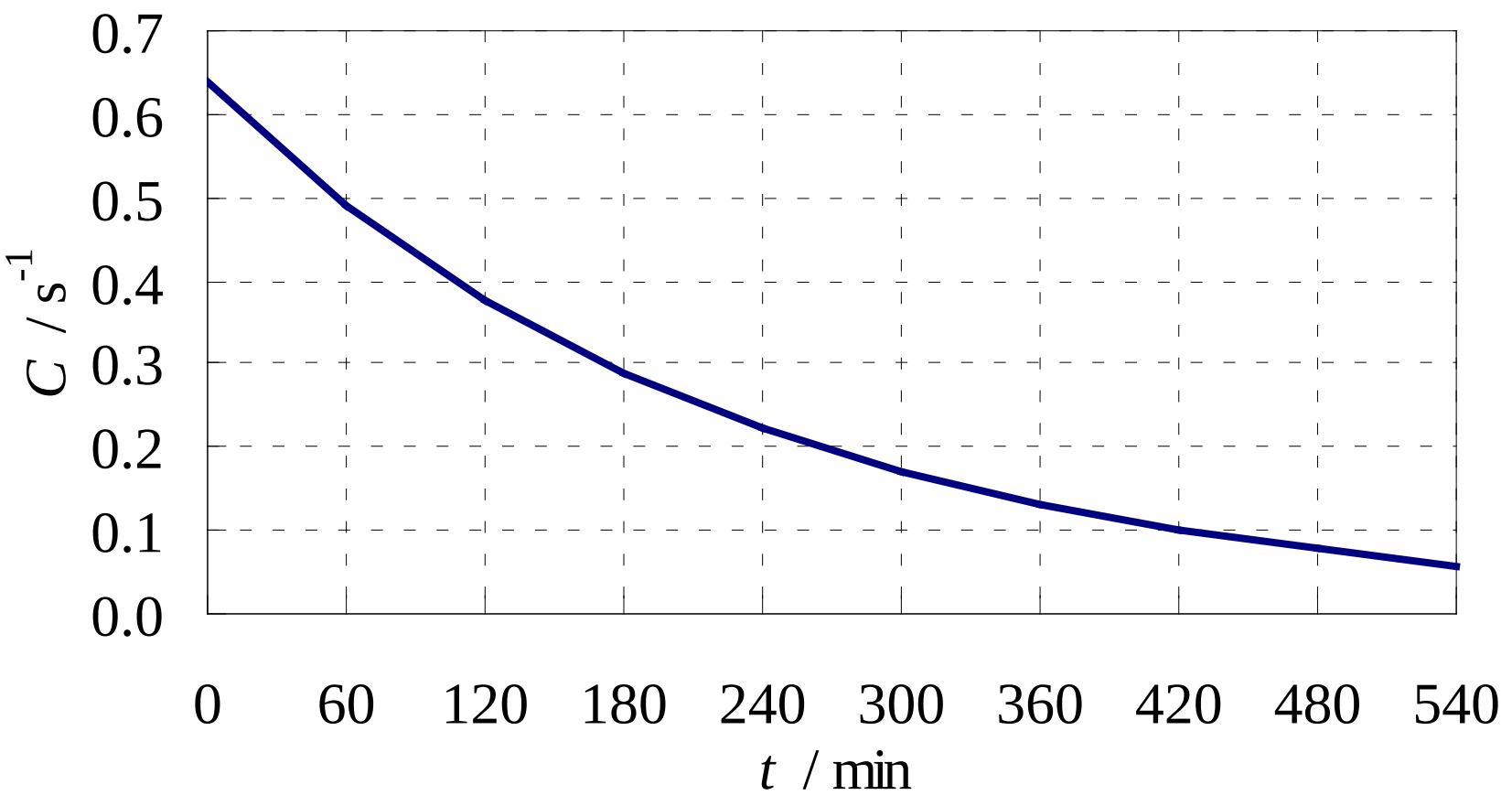

**Figure 5.** Expected $^{31}Si$ count rate of a 6 g AVO28 sample; $t = 0$ min corresponds to the irradiation end.

The net counts, $n_c$, stored in a hypothetical spectrometry sequence starting immediately after irradiation and consisting of nine consecutive counts having a counting window of 60 min are shown in the upper graph of figure 6. The count uncertainties,

$$u(n_c) = \sqrt{n_c + 2\,n_b}\,, \qquad (7)$$

depend on the expected background count, $n_b$, at the energy bins of the multichannel analyzer calibrated to collect the $^{31}Si$ $\gamma$-photons [18]. Since the expected operating conditions of the detection system when measuring the AVO28 sample will be close to the operating conditions at the end of the second sequence of this test, the associated 0.07 $s^{-1}$ background count rate is, though rough, a realistic estimate of the expected background count rate. The expected uncertainties, shown in the lower graph of figure 6, are calculated according to the equation (7), where the $n_b$ value refers to a counting window of 60 min.

Detectors having a 1266.1 keV peak efficiency higher than the Ge detector used in this feasibility study ($\varepsilon = 0.029$ at 1 cm) will deliver higher count rates and lower count uncertainties. A well-type Ge detector, a ultra-high efficiency Ge detector (150% relative efficiency) or a large scintillation detector could be fit of purpose. For example, the 0.152 efficiency at 1266.1 keV of the 350 $cm^3$ ORTEC well detector would increase the expected AVO28 count rate after irradiation to about 5.24 $s^{-1}$.

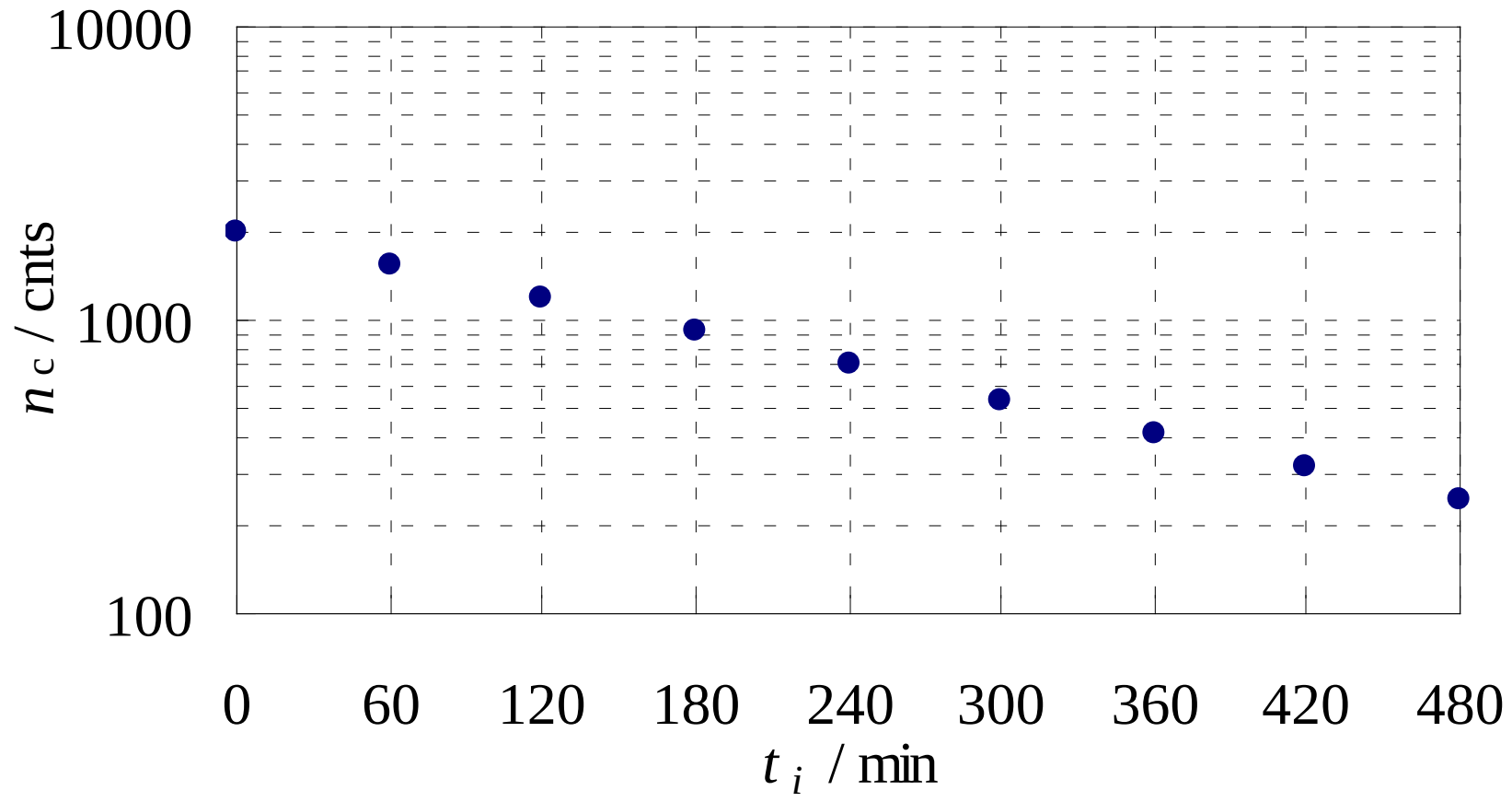

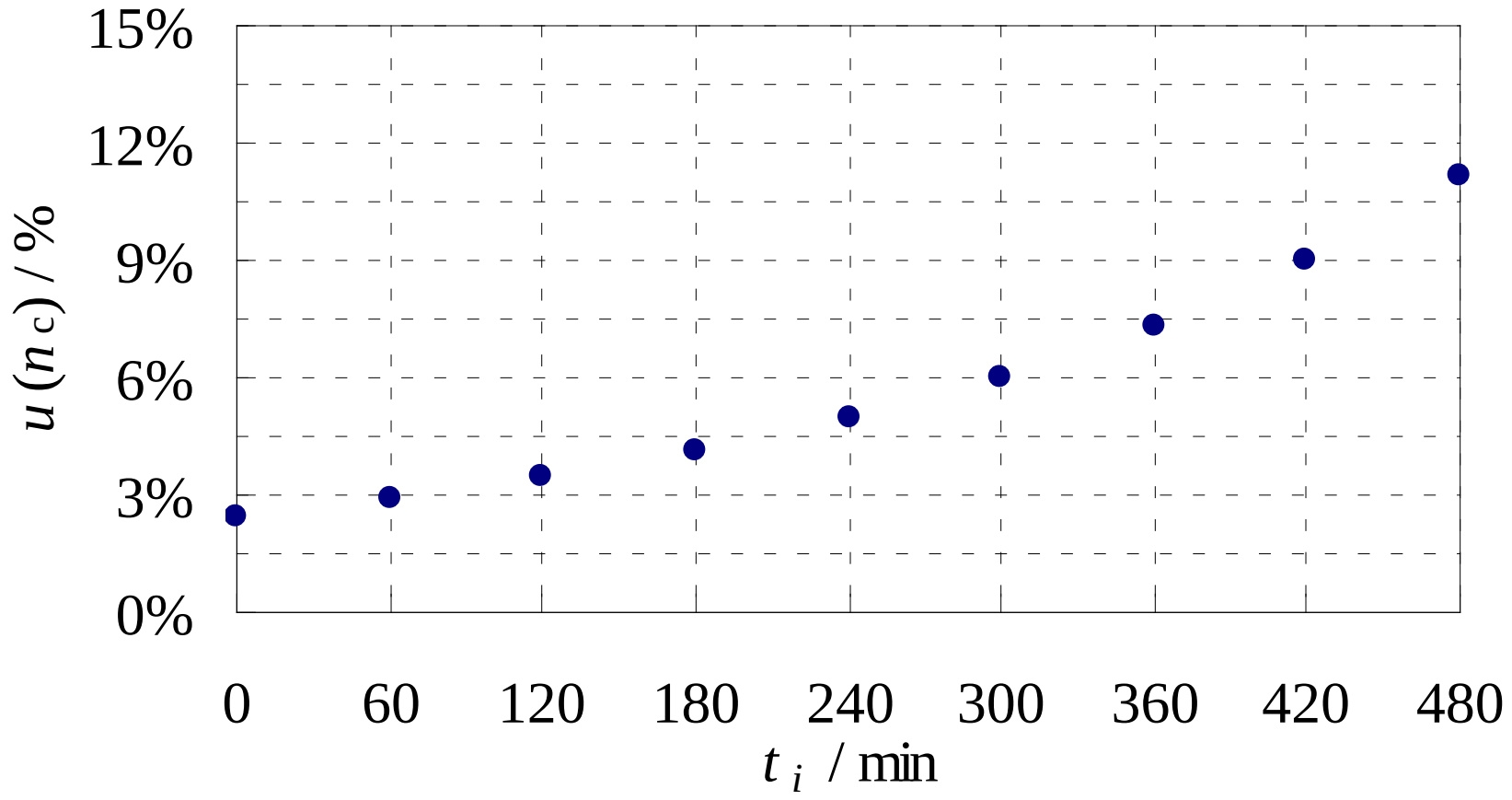

**Figure 6.** Expected consecutive net counts of the 1266.1 keV peak for a 6 g AVO28 sample (upper graph) and the associated relative standard uncertainties assuming a background count rate of about 0.07 $s^{-1}$ (lower graph); $t_i$ = 0 min corresponds to the irradiation end.

## 4. Expected uncertainty

The measurement model (5) was used to evaluate the expected uncertainty [19] of the result that will be obtained by the application of the proposed method to a cylindrical 6 g AVO28 sample when using an identical cylindrical 6 g WASO04 sample as a standard. The budget is referred to a 3 h neutron irradiation at the central thimble of a 250 kW TRIGA Mark II reactor and to the counting of the AVO28 and WASO04 samples in two spectrometry sequences started at about 3 h and 26 h after irradiation, respectively, at 1 cm from a Ge detector having a 50% relative efficiency.

The count rates, $C_{\mathrm{STD}}$ and $C_{\mathrm{AVO28}}$, are expected to be mainly affected by the uncertainties of the net counts $n_{\mathrm{c}}$ in the relevant spectrometry sequences. The uncertainties of the $n_{\mathrm{c}}$ values recorded in the second spectrometry sequence of the WASO04 sample and the uncertainties of the $n_{\mathrm{c}}$ values obtained in the extrapolated spectrometry sequence of the AVO28 sample were used to compute the expected relative combined standard uncertainty of $C_{\mathrm{STD}}$ and $C_{\mathrm{AVO28}}$. They were found to be $3 \times 10^{-3}$ and $2.5 \times 10^{-2}$, respectively. Only the $n_{\mathrm{c}}$ values whose relative standard uncertainty is less than 10% have been considered; besides, this is necessary also to avoid biased result.

The uncertainty of the $t_{1/2}$ value of $^{31}$Si is the major contribution to the uncertainty of the correction factor $\kappa_{\mathrm{td}}$. The decay time between the starts of the two counting sequences, $t_{\mathrm{d\,STD}} - t_{\mathrm{d\,AVO28}}$, is expected about 9 $t_{1/2}$. By propagating the uncertainty linearly, the uncertainty of the $t_{1/2}$ value obtained in this study induces a relative combined standard uncertainty of the $\kappa_{\mathrm{td}}$ value of about $2 \times 10^{-2}$.

When using a high-purity natural-Si as a standard, $x(^{30}\mathrm{Si}_{\mathrm{STD}})$ and $A_{\mathrm{r\,STD}}$ are known to within negligible relative combined standard uncertainties, about $2 \times 10^{-3}$ and $5 \times 10^{-6}$, respectively [12, 20]. Additionally, since $x(^{29}\mathrm{Si}_{\mathrm{AVO28}})$ and $x(^{30}\mathrm{Si}_{\mathrm{AVO28}})$ are at $10^{-5}$ and $10^{-6}$ level, respectively, $A_{\mathrm{r\,AVO28}} \cong A_{\mathrm{r}}(^{28}\mathrm{Si})$ to within $10^{-5}$ and $A_{\mathrm{r}}(^{28}\mathrm{Si})$ is known to within negligible relative combined standard uncertainty, $6.9 \times 10^{-11}$ [21].

The gravimetric measurement of the masses, $m_{\mathrm{STD}}$ and $m_{\mathrm{AVO28}}$, can be carried out with a relative combined standard uncertainty less than $2 \times 10^{-4}$, when weighing samples having a mass greater than 1 g.

The application of the analytical expression to calculate the neutron self-shielding in a cylindrical sample [22] and the characterisation of the Ge detector showed that a 0.5 mm tolerance of the sample length and diameter corresponds to negligible deviations, about $2 \times 10^{-4}$, from a unit value of the correction factors $\kappa_{\mathrm{ss}}$, $\kappa_{\mathrm{sa}}$ and $\kappa_{\mathrm{g}}$. Additionally, assuming a 0.5 mm tolerance of the counting position, $\kappa_{\varepsilon} = 1$ with a relative combined standard uncertainty of $1 \times 10^{-2}$.

As regards as the difference between the energy spectra of the neutrons irradiating the standard and AVO28 samples, under the reasonable assumption that the flux of fission neutrons is negligible, the $E^{-1/2}$ dependence of the cross section of the reaction $^{30}$Si(n,γ)$^{31}$Si allows the reaction rate per atom to be expressed by the Høgdahl convention [23]. Since only the difference of the energy-spectra amplitudes is suspected to be significant, $\kappa_{\mathrm{R}} = \Phi_{\mathrm{th\,STD}} / \Phi_{\mathrm{th\,AVO28}}$, where $\Phi_{\mathrm{th\,STD}}$ and $\Phi_{\mathrm{th\,AVO28}}$ are the thermal-neutron fluxes irradiating the two samples. Thus, the $\kappa_{\mathrm{R}}$ value can be estimated via flux monitors located near the samples. A preliminary test carried out with Co-Al wires showed that a correction can be applied with a relative combined standard uncertainty of about $3 \times 10^{-3}$.

The expected uncertainty budget of the measurement of the $^{30}$Si mole fraction of the AVO28 material by INAA (based on the information collected in this study) is given in table 2. The overriding contributors to the uncertainty budget are $\kappa_{\varepsilon}$, $C_{\mathrm{AVO28}}$ and $\kappa_{\mathrm{td}}$. An overall relative combined standard uncertainty of about 4% is conservatively evaluated.

| Quantity $X_i$ | Relative standard uncertainty $u(x_i)$ | Index % |
|---|---|---|
| $\kappa_R$ | $3.0 \times 10^{-3}$ | 0.8 |
| $\kappa_{td}$ | $2.0 \times 10^{-2}$ | 34.9 |
| $\kappa_\varepsilon$ | $1.0 \times 10^{-2}$ | 8.7 |
| $\kappa_{ss}$ | $2.0 \times 10^{-4}$ | 0.0 |
| $\kappa_{sa}$ | $2.0 \times 10^{-4}$ | 0.0 |
| $\kappa_g$ | $2.0 \times 10^{-4}$ | 0.0 |
| $C_{AVO28}$ | $2.5 \times 10^{-2}$ | 54.5 |
| $C_{STD}$ | $3.0 \times 10^{-3}$ | 0.8 |
| $x(^{30}Si_{STD})$ | $2.0 \times 10^{-3}$ | 0.3 |
| $m_{STD}$ | $2.0 \times 10^{-4}$ | 0.0 |
| $m_{AVO28}$ | $2.0 \times 10^{-4}$ | 0.0 |
| $A_{rAVO28}$ | $1.0 \times 10^{-5}$ | 0.0 |
| $A_{rSTD}$ | $5.0 \times 10^{-6}$ | 0.0 |
| $Y$ | $u_c(y)$ | |
| $x(^{30}Si_{AVO28})$ | $3.4 \times 10^{-2}$ | 100.0 |

**Table 2.** Expected uncertainty budget of the INAA measurement result of the $^{30}$Si mole fraction of AVO28 material. The input quantities $x_i$ are given in section 2.1. The relative standard uncertainties, $u(x_i)$, have unit coverage factor. The Index column gives the relative contributions of $u(x_i)$ to the total relative combined standard uncertainty, $u_c(y)$, of $x(^{30}Si_{AVO28})$.

## 5. Conclusions

We proved that a relative method based on neutron activation analysis can assess the $^{30}$Si mole fraction of a silicon crystal highly enriched in $^{28}$Si. A preliminary proof of concept using natural silicon allowed us to conclude that the analysis of the enriched material might achieve a 4% relative combined standard uncertainty. Therefore, this measurement delivers useful and independent information on the isotopic composition of the material used for the measurement of the Avogadro constant.

## Acknowledgments

This work was jointly funded by the European Metrology Research Programme (EMRP) participating countries within the European Association of National Metrology Institutes (EURAMET) and the European Union. The authors whish to thank John Bennett and Robert Vocke for their fruitful comments and the revision of an initial draft of this paper. Moreover, they are grateful to M. Santiano for the machining the counting containers and M. Clemenza for providing the efficiency data of the well-type detector.